\documentclass[12pt]{article}  

\usepackage{graphicx}
\usepackage{amsfonts}
\usepackage{amsmath}
\usepackage{amssymb}

\makeatletter
\def\vereq#1#2{\lower3pt\vbox{\baselineskip1.5pt \lineskip1.5pt
\ialign{$\m@th#1\hfill##\hfil$\crcr#2\crcr\sim\crcr}}}
\makeatother

\begin{document}

\begin{titlepage}

\begin{flushright}
\end{flushright}

\vskip 3cm

\begin{center}
{\Large \bf  $ b \rightarrow s \gamma$ in a Calculable Model \\
             of Electroweak Symmetry Breaking}

\vskip 1.0cm

{\bf
Riccardo Barbieri$^{a}$,
Giacomo Cacciapaglia$^{a}$,
Alessandro Romito$^{a}$
}

\vskip 0.5cm

$^a$ {\it Scuola Normale Superiore and INFN, Piazza dei Cavalieri 7, 
                 I-56126 Pisa, Italy}\\

\vskip 1.0cm

\abstract{In a recently proposed extension of the Standard Model with a compact extra dimension of size $R$ we compute the leading corrections, of relative order $(m_{t} R)^2$, to the Branching Ratio $\mathcal{BR}(B \rightarrow X_{s} \gamma)$. For $R = (370 \pm 80 GeV)^{-1}$ the branching ratio is increased in a significant way relative to the SM, although not at the level of being inconsistent with present measurements.}

\end{center}
\end{titlepage}

\section{Introduction}

To entertain the possibility that there exist extra space dimensions gives a framework for new ideas and interesting intellectual exercises. 
For many a strong motivation in this direction has been and is provided by string theory. 
On the other hand, a revival of interest about extra dimensions has arisen from the relatively more recent realization that their size, if compact, need not to be so small as usually thought \cite{anton}. 
In turn, this has led to a discussion which goes under the name of phenomenology of extra dimensions: the existence of extra dimensions could be amenable to experimental verification. 
Collider physics \cite{giudice}, modifications of gravity at variuos distances \cite{ADD}  or even neutrino physics have been considered in this context.

In our view this discussion, although interesting, generally suffers from a serious drawback: the lack of a quantitative or semi-quantitative connection between a known physical scale and the scale at which these new phenomena could take place. 
This lack undermines the relevance of any concrete phenomenological study.
As an example, even the otherwise appealing idea \cite{ADD} that the quantum gravity scale $M$, perhaps the shortest possible scale in physics, could be related to the Fermi scale can be hardly evaluated on phenomenological grounds because of this problem.

A specific proposal which does not suffer from this drawback was made in Ref \cite{BHN}.
There one considers a supersymmetric theory in 5 dimensions with the Standard Model gauge group and one 5D hypermultiplet for every matter or Higgs gauge multiplet of the SM. 
Given this starting point, the model is defined by taking the physical space of the 5th dimension as the only orbifold which produces the SM particles as massless modes.
Remarkably, the residual symmetry after the orbifold compactification includes a $N=2$ supersymmetry with local transformation parameters which also satisfy a precise orbifold condition \cite{BHNinp}.
As a consequence, the most general Lagrangian compatible with this residual symmetry is highly restricted, leading to the calculability of various physical observables. 
An example explicitly discussed in Ref \cite{BHN} is the Higgs mass, which is directly related to the compactification radius $R$.
Other cases include the $g-2$ of the muon, the parameter $\epsilon_{3}$, or $S$, in electroweak precision physics or the branching ratio for the decay of the Higgs in two photons. 

In this paper we describe the calculation of the $b \rightarrow s \gamma$ decay with the inclusion of the leading corrections to the SM amplitude of relative order $(m_{t} R)^2$, where $m_{t}$ is the top quark mass.

\section{The model \cite{BHN,BHNinp}}

The salient features of the model are readily summarized.
The gauge group is $SU(3)\times SU(2)\times U(1)$ and the hypermultiplets can be denoted as $Q$, $U$, $D$, $L$, $E$, $H$, borrowing the usual notation from the SM.
We remind that a hypermultiplet contains, from a 4D point of view, a chiral supermultiplet an another supermultiplet with conjugate quantum numbers, labelled by a ``c''.
From 5D, the orbifold compactification that leads to the SM particles as the only massless states, before Electroweak Symmetry Breaking, is $S^{1}/\mathbb{Z}_{2}\times \mathbb{Z}'_{2}$.
Under the two $\mathbb{Z}_{2}$ parities, which act at $y=0$ and $y=\pi R/2$ respectively,  all the fields have definite transformation properties, given in Table \ref{tabella} together with the corresponding eigenfunctions and the spectrum of every tower of Kaluza-Klein states.
Only the towers corresponding to $(+,+)$ contain the zero modes.

\begin{table}[tb]
\begin{center}
\begin{tabular}{|c|c|c|c|}
\hline
$A_{\mu}$, $\psi_{M}$, $\phi_{H}$ & $(+, +)$ & $\displaystyle \cos{\frac{2n}{R} y}$ & $\displaystyle \frac{2n}{R}$, n = 0, 1, 2, 3...\\
\hline
$\lambda$, $\phi_{M}$, $\psi_{H}$ & $(+, -)$ & $\displaystyle \cos{\frac{2n-1}{R} y}$ & $\displaystyle \frac{2n-1}{R}$, n = 1, 2, 3...  \\
\hline
$\psi_{\Sigma}$, $\phi_{M}^{c\dagger}$, $\psi_{H}^{c\dagger}$ & $(-, +)$ & $\displaystyle \sin{\frac{2n-1}{R} y}$ & $\displaystyle \frac{2n-1}{R}$, n = 1, 2, 3...\\
\hline
$\phi_{\Sigma}$, $\psi_{M}^{c\dagger}$, $\phi_{H}^{c\dagger}$ & $(-, -)$  & $\displaystyle \sin{\frac{2n}{R} y}$ & $\displaystyle \frac{2n}{R}$, n = 1, 2, 3...\\
\hline
\end{tabular}
\end{center}
\caption{Matter, Higgs and gauge fields content of the theory with their orbifolding properties} \label{tabella}
\end{table}

In spite of the manifestly non supersymmetric spectrum, after the orbifold projection, the theory is still invariant under suitable supersymmetric transformations.
From the 4D viewpoint, the theory possesses 2 local supersymmetries with tranformation parameters $(\xi_{1}(x,y), \xi_{2}(x,y))$ behaving as $(+, -)$ and $(-, +)$ under $\mathbb{Z}_{2} \times \mathbb{Z}'_{2}$.
In turn, this invariance, together with the gauge symmetry, fixes the general form of the 5D Lagrangian to:

\begin{equation} \label{genlagr}
\mathcal{L} (x, y) = \mathcal{L}_{5} + \delta(y) \mathcal{L}_{4} + \delta(y - \pi R/2) \mathcal{L}'_{4}.
\end{equation}
$\mathcal{L}_{5}$ is $N=1$ supersymmetric in 5D whereas $ \mathcal{L}_{4}$ and $\mathcal{L}'_{4}$ are 4D lagrangians invariant under $N=1$ supersymmetries induced by $( \xi_{1}(x,0), 0)$ and $(0, \xi_{2}(x,\pi R/2))$ respectively.

Most relevant to the present discussion are the supersymmetric Yukawa couplings of the Higgs to the up, $\lambda_{U} \hat{Q} \hat{U} \hat{H}$, and to the down quarks, $\lambda_{D} \hat{Q}' \hat{D}' \hat{H}'_{c}$.
These two couplings are localized at $y=0$ and $y=\pi R/2$ respectively.
Since the supersymmetric transformations at $y=0$ and $y=\pi R/2$ are different, this explains the prime indices on the $N=1$ chiral multiplets in the $\lambda_{D}$-couplings.
It is:

$$
\begin{array}{ccc}
\hat{Q} = \left( \begin{array}{c} \psi_{Q} \\ \phi_{Q} \end{array} \right) & \hat{U} = \left( \begin{array}{c} \psi_{U} \\ \phi_{U} \end{array} \right) & \hat{H} = \left( \begin{array}{c} \psi_{H} \\ \phi_{H} \end{array} \right) \\
\hat{Q}' = \left( \begin{array}{c} \psi_{Q} \\ \phi^{c \dagger}_{Q} \end{array} \right) & \hat{D}' = \left( \begin{array}{c} \psi_{D} \\ \phi^{c \dagger}_{D} \end{array} \right) & \hat{H}'_{c} = \left( \begin{array}{c} \psi^{c}_{H} \\ \phi^{\dagger}_{H} \end{array} \right) 
\end{array}
$$
Note in particular that both the scalar components of $\hat{H}$ and $\hat{H}'_{c}$ involve $\phi_{H}$ which is the only scalar with a zero mode, i.e. the standard Higgs field.
As a consequence of radiative corrections, $\phi_{H}$ gets a vev which occurs at the Fermi scale if $1/R = 370 \pm 80 GeV$, with an estimated error due to the effects of kinetic terms at the fixed points, $y=0$ and $y=\pi R/2$, and to higher order corrections \cite{BHN,BHNinp}.

After mode expansion, the Yukawa couplings to the charged Higgs have the explicit form:

\begin{equation} \label{Lint}
\begin{array}{rcl}
\mathcal{L}_{int} & = & \displaystyle  -\frac{f_{t}}{\sqrt{2}} \displaystyle \left( V_{tb} \psi_{b} + V_{ts} \psi_{s}\right) \sum_{m,n=0}^{\infty} \eta_{n}^{\psi} \eta_{m}^{\psi}  \psi_{U,n} \phi_{H,m}  + h.c.  \\
& & + \displaystyle \frac{f_{b}}{\sqrt{2}} \displaystyle V_{tb} \psi_{b^{c}} \sum_{m,n=0}^{\infty} \xi_{n}^{\psi} \xi_{m}^{\psi}  \psi_{ T,n} \phi^{\dagger}_{H,m} + h.c.  \\
& & \displaystyle  - \frac{f_{t}}{\sqrt{2}} \left( V_{tb} \psi_{b} + V_{ts} \psi_{s}\right)  \displaystyle \sum_{m,n=1}^{\infty} \eta_{n}^{\phi} \eta_{m}^{\phi}  \psi_{H,m} \phi_{U,n} + h.c. \\
& & + \displaystyle \frac{f_{b}}{\sqrt{2}} V_{tb} \psi_{b^{c}}  \displaystyle \sum_{m,n=1}^{\infty} \xi_{n}^{\phi} \xi_{m}^{\phi}  \psi^{c}_{H,m} \phi_{T,n} + h.c.
\end{array}
\end{equation}
where:

\begin{equation}
\begin{array}{rcl}
\eta^{\psi}_{n} & = & \left( \frac{1}{\sqrt{2}} \right)^{\delta_{n,0}} \\
\eta^{\phi}_{n} & = & 1\\
\xi^{\psi}_{n} & = & \left( \frac{1}{\sqrt{2}} \right)^{\delta_{n,0}} (-1)^{n}\\
\xi^{\phi}_{n} & = & (-1)^{n+1}
\end{array}
\end{equation}
are the wave functions of the different modes at $y=0$ or $y=\pi R/2$.
As usual we have diagonalized the Yukawa matrices $\lambda_{U}$ and $\lambda_{D}$ in the coupling to the neutral Higgs not shown in (\ref{Lint}), while we have kept the CKM matrix $V$ in the couplings to the charged Higgs. 
$\psi_{b}$, $\psi_{s}$ and $\psi_{b^{c}}$ are the left and right components of the $b$ and $s$ quarks, i.e. only the corresponding zero modes. 
As in the SM calculation of the $b \rightarrow s \gamma$ amplitude, all quark masses are neglected except for the top and the bottom ones.

The gauge couplings to the charged $W$-boson and its tower would also contain the flavour changing CKM matrix.
We do not show them, however, since we shall only be computing the corrections to the SM amplitude of relative order $(m_{t} R)^2$.
Loops with $W$-exchanges, either zero or higher modes, would only correct the SM result by terms of relative order $(m_{W} R\; m_{t} R)^2$, that we neglect.

\section{The diagonalized Yukawa couplings}

The non conservation of momentum in the 5th direction gives also rise to mass mixings in the top and stop towers, which is convenient to diagonalize at the very beginning.
In the fermion sector, we denote the two Weyl components of the Dirac mass eigenstates by $(\psi_{0}, \psi^{c}_{0})$, $(\psi_{n}^{+}, \psi_{n}^{c+})$ and $(\psi_{n}^{-}, \psi_{n}^{c-})$ with mass eigenvalues $m_{0}$, $m_{n}^{\pm}$, $n=1,2,...$, while in the relevant stop sector we use $\Phi_{n}^{+}$, $\Phi_{n}^{-}$ with eigenvalues $\tilde{m}_{n}^{+}$, $\tilde{m}_{n}^{-}$, $n = 1, 2, ...$.
The diagonalization is made in App \ref{mass} for both cases to any order in 
$$
\epsilon = \displaystyle \frac{f_{t}}{2^{3/2}} v R  \approx m_{t} R
$$
To the order of interest for $b \rightarrow s \gamma$, it is:

\begin{equation}
m_{0} = \epsilon - \frac{\pi^2}{12} \epsilon^{3}, \quad m_{n}^{\pm} = \epsilon \pm 2n, \quad \tilde{m}_{n}^{\pm} = 2n-1 \pm \epsilon
\end{equation}
in unit of $1/R$ and

\begin{eqnarray}
\psi_{T,0} & = & \displaystyle \psi_{0} \left( 1 - \frac{\pi^2}{24} \epsilon^{2}\right) + \epsilon \sum_{n=1}^{\infty} \frac{1}{2n} ( \psi_{n}^{+} - \psi_{n}^{-} ) \label{psit0} \\ 
\displaystyle \psi_{T,l} & = & \displaystyle  \frac{1}{\sqrt{2}} \left[ \left( 1 + \frac{\epsilon}{2(2l)} \right) \psi^{+}_{l} + \left( 1 - \frac{\epsilon}{2(2l)} \right) \psi^{-}_{l} \right]  + \epsilon \frac{1}{\sqrt{2}} \sum_{n=1}^{\infty} d_{nl} (\psi_{n}^{+} - \psi_{n}^{-} ) - \epsilon^{2}  \frac{\sqrt{2}}{(2l)^{2}} \psi_{0} \label{psitl} \\
\displaystyle \psi^{c}_{U,l} & = &  \displaystyle \frac{1}{\sqrt{2}} \left[ \left( 1 - \frac{\epsilon}{2(2l)} \right) \psi^{+}_{l} - \left( 1 + \frac{\epsilon}{2(2l)} \right) \psi^{-}_{l} \right]  - \epsilon \frac{1}{\sqrt{2}} \sum_{n=1}^{\infty} d_{ln} (\psi_{n}^{+} + \psi_{n}^{-} ) - \epsilon \frac{\sqrt{2}}{(2l)} \psi_{0} \label{psicul} \\
\phi_{U,l} & = &  \displaystyle \frac{1}{\sqrt{2}} \left( (1 + \frac{\epsilon}{2(2l-1)} ) \Phi^{+}_{l} + (1 - \frac{\epsilon}{2(2l-1)} ) \Phi^{-}_{l} \right) + \frac{1}{\sqrt{2}} \epsilon \sum_{n=1}^{\infty} c_{nl} \left( \Phi^{+}_{n} - \Phi^{-}_{n} \right) \label{phiul}\\
\phi_{T,l}^{c} & = & \displaystyle  \frac{1}{\sqrt{2}} \left( (1 + \frac{\epsilon}{2(2l-1)} ) \Phi^{-}_{l} - (1 - \frac{\epsilon}{2(2l-1)} ) \Phi^{+}_{l} \right) + \frac{1}{\sqrt{2}} \epsilon \sum_{n=1}^{\infty}c_{ln} \left( \Phi^{+}_{n} + \Phi^{-}_{n} \right) \label{phitcl}
\end{eqnarray}
where $d_{ln} = \frac{2(2l)}{(2l)^2 - (2n)^2}$ and $c_{ln} = \frac{2(2l-1)}{(2l-1)^2 - (2n-1)^2}$ for $l \neq n$ and zero for $l=n$.
Since the fermion mass matrix is symmetric, identical expressions to (\ref{psit0}, \ref{psitl}, \ref{psicul}) hold for $(\psi_{U,0}, \psi_{U,l}, \psi^{c}_{T,l})$ with $(\psi_{0}, \psi^{\pm}_{l})$ replaced by $(\psi^{c}_{0}, \psi^{c \pm}_{l})$.

Equations (\ref{Lint}) and (\ref{psit0}-\ref{phitcl}) allow to write the Yukawa couplings to the mass eigenstates:

\begin{equation}
\begin{array}{rcl}
 \mathcal{L}_{int} & = & \displaystyle  -\frac{f_{t}}{\sqrt{2}} (V_{tb} \psi_{b} + V_{ts} \psi_{s} ) \left( \displaystyle \sum_{m=0}^{\infty} \eta^{\psi}_{m} \phi_{H,m} \right) \left(\displaystyle \sum_{n=-\infty}^{\infty} \mu^{c}_{n} \psi_{n}^{c} \right) + h.c.\\
 & & \displaystyle  + \frac{f_{b}}{\sqrt{2}} V_{tb} \psi_{b^{c}} \left( \displaystyle \sum_{m=0}^{\infty} \xi^{\psi}_{m} \phi^{\dagger}_{H,m} \right) \left( \displaystyle \sum_{n=-\infty}^{\infty}  \mu_{n} \psi_{n}\right) + h.c.\\
 &  & \displaystyle  -\frac{f_{t}}{\sqrt{2}} (V_{tb} \psi_{b} + V_{ts} \psi_{s} ) \left( \displaystyle \sum_{m=1}^{\infty} \eta^{\phi}_{m} \psi_{H,m} \right)  \left( \displaystyle \sum_{n=-\infty}^{\infty}  \nu^{c}_{n}  \Phi_{n}   \right) + h.c.\\
 & & \displaystyle  + \frac{f_{b}}{\sqrt{2}} V_{tb} \psi_{b^{c}} \left( \displaystyle \sum_{m=1}^{\infty} \xi^{\phi}_{m} \psi^{c}_{H,m} \right)  \left( \displaystyle \sum_{n=-\infty}^{\infty} \nu_{n}  \Phi_{n}  \right) + h.c.
\end{array} \end{equation}
where we have set

\begin{equation} \psi_{r} = \left\{ \begin{array}{cc}
r > 0 & \psi_{r}^{+}\\
r = 0 & \psi_{0} \\
r < 0 & \psi_{-r}^{-}
\end{array} \right.
\end{equation}
and similarly for the other fields.
It is:
\begin{equation}
\mu^{c}_{r} = \left\{ \begin{array}{cc}
r > 0 & \frac{1}{\sqrt{2}} \\
r = 0 & \frac{1}{\sqrt{2}} \left( 1 - \frac{\pi^2}{8} \epsilon^2 \right)\\
r < 0 & \frac{1}{\sqrt{2}} 
\end{array} \right. \qquad \mu_{r} = \left\{ 
\begin{array}{cc}
r > 0 & \frac{(-1)^r}{\sqrt{2}} \\
r = 0 & \frac{1}{\sqrt{2}} \\
r < 0 & \frac{(-1)^r}{\sqrt{2}} 
\end{array} \right.
\end{equation}

\begin{equation}
\nu_{r}^{c} = \left\{ \begin{array}{cc}
r > 0 & \frac{1}{\sqrt{2}}\\
r = 0 & 0\\
r < 0 & \frac{1}{\sqrt{2}}
\end{array} \right. \qquad \nu_{r} = \left\{ 
\begin{array}{cc}
r > 0 & - \frac{(-1)^{r+1}}{\sqrt{2}}\\
r = 0 & 0\\
r < 0 & \frac{(-1)^{r+1}}{\sqrt{2}}
\end{array} \right.
\end{equation}
whereas, for the masses of the top and stop towers respectively, we have:

\begin{equation}
m_{r} = \left\{ \begin{array}{cc}
r > 0 & \epsilon + 2r \\
r = 0 & \epsilon - \frac{\pi^2}{12} \epsilon^3 \\
r < 0 & \epsilon + 2r 
\end{array} \right. \qquad \tilde{m}_{r} = \left\{ \begin{array}{cc}
r > 0 & 2r-1 + \epsilon\\
r = 0 & 0\\
r < 0 & -2r-1 - \epsilon
\end{array} \right.
\end{equation}

\section{Calculations} \label{calculation}

\begin{figure}[bt]
\begin{center}
\includegraphics[width=12cm]{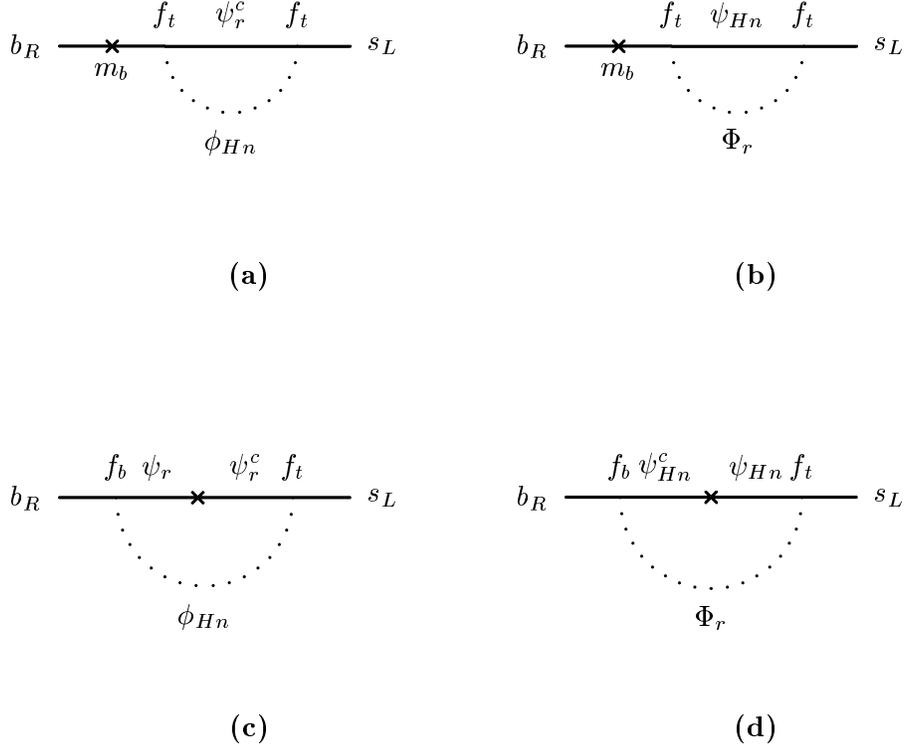}
\end{center}
\caption{Graphs contributing to the $b \rightarrow s \gamma$ effective operator. The external photon has to be attached to any of the internal lines.} \label{grafici}
\end{figure}

In the approximation we are working, four diagrams contribute to the $b \rightarrow s \gamma$ amplitude, shown in Figure \ref{grafici}.
A photon attached to any of the internal lines is left understood.
Note that in the internal lines all the KK are propagating, with no conservation of momentum in the 5th direction at the vertices of the Yukawa couplings, which are localized on the branes.
This is similar to the case of the top loop correction to the Higgs mass described in Ref \cite{BHN} .
A novelty in the present case is that diagrams (1c) and (1d) involve both couplings at different $y$ locations.

As usual, for every diagram we define the dimensionless coefficients $C$ appearing in their corresponding contribution to the effective Lagrangian:

\begin{equation}
\mathcal{L}_{eff} = - \frac{G_{F}}{\sqrt{2}} V_{tb} V^{*}_{ts} \frac{e}{4\pi^{2}} m_{b} (\overline{s}_{L} \sigma_{\mu \nu} F^{\mu \nu} b_{R}) C
\end{equation}

For every diagram in Fig \ref{grafici} and for individual intermediate states, a straightforward calculation gives:

\begin{eqnarray}
C_{nr}^{a} & = & 4 \epsilon^{2} \int_{0}^{\infty} 2 q^{3} dq \int_{0}^{1} dx  \frac{P^{a} (x) }{[q^{2} + (2n)^{2} x + (m_{r})^{2} (1-x)]^{3}}\left( \eta_{n}^{\psi}\right)^{2} \left( \mu_{r}^{c} \right)^2 \label{grapha}\\
C_{nr}^{b} & = & - 4 \epsilon^{2} \int_{0}^{\infty} 2 q^{3} dq \int_{0}^{1} dx  \frac{P^{b}(x) }{[q^{2} + (2n-1)^{2} x + (\tilde{m}_{r})^{2} (1-x)]^{3}} \left( \eta_{n}^{\phi}  \right)^{2} \left( \nu_{r}^{c} \right)^2\\
C_{nr}^{c} & = & 4 \epsilon  \int_{0}^{\infty} 2 q^{3} dq \int_{0}^{1} dx  \frac{P^{c}(x)}{[q^{2} + x (2n)^{2} + (1-x) (m_{r})^{2} ]^{3}} m_{r} \eta^{\psi}_{n}  \xi^{\psi}_{n} \mu_{r}^{c} \mu_{r} \\
C_{nr}^{d} & = & - 4 \epsilon  \int_{0}^{\infty} 2 q^{3} dq \int_{0}^{1} dx  \frac{P^{d}(x)}{[q^{2} + x (\tilde{m}_{r})^{2} + (1-x) (2n-1)^{2} ]^{3}} (2n-1) \eta^{\phi}_{n}  \xi^{\phi}_{n} \nu_{r}^{c} \nu_{r} \label{graphd}
\end{eqnarray}
where:
$$
P^{a}(x) = P^{b}(x) = -\frac{1}{2} \left( \frac{2}{3} x (1-x)^2 + x^{2} (1-x)\right)
$$
$$
P^{c} (x) = \frac{2}{3} (1-x)^2 + x (1-x)
$$
$$
P^{d}(x) = (1-x)^2 + \frac{2}{3} x (1-x)
$$

\subsection{Diagrams (1a) and (1b)}

In the calculation of diagram (1a) it is convenient to separate the contributions with at least one zero mode.
Therefore we define:

\begin{equation}
C^{a} \equiv C^{a}_{K0} + C^{a}_{0K} + C^{a}_{KK}
\end{equation}
where:

\begin{equation}
C^{a}_{K0} = \displaystyle \sum_{n=0}^{\infty} C^{a}_{n0} , \quad C^{a}_{0K} = \displaystyle \sum_{r\neq 0} C^{a}_{0r}, \quad C^{a}_{KK} = \sum_{n\neq0, r\neq0} C^{a}_{nr}
\end{equation}

A straightforward calculation gives:

\begin{equation}
C^{a}_{K0} = -\frac{1}{9} + \frac{\pi^2}{864} \epsilon^2
\end{equation}

\begin{equation}
C^{a}_{0K}  = - \frac{\pi^2}{108} \epsilon^2
\end{equation}

Although the individual contributions to $C^{a}_{KK}$ are finite, the double sum would lead to a divergent result.
Convergence arises only if the supersymmetric diagram (1b), which does not have zero modes in its internal lines, is added to (1a) and the momentum integration is performed after the summation over the KK modes.
Defining:

\begin{equation}
C^{b} = \sum_{n,r} C^{b}_{nr}
\end{equation}
it is:

\begin{equation} \label{cabkk}
C^{a}_{KK} + C^{b}  = 4 \epsilon^2 \displaystyle \int_{0}^{\infty} 2 q^3 dq F_{ab}(q) = 0.215 \epsilon^2
\end{equation}
where: 

\begin{eqnarray*}
F_{ab} (q) & = & - \frac{5}{24} \int_{q^{2}}^{\infty} dy^{2} \left[ (f'_{1})^{2} - (f'_{2})^{2} \right] \\
f_{1} (y) & = & \sum_{n=1}^{\infty} \frac{1}{y^2 + (2n)^{2}} = \frac{\pi y  \coth{\frac{\pi}{2} y} -2}{4 y^2}\\
f_{2} (y) & = & \sum_{n=1}^{\infty} \frac{1}{y^2 + (2n-1)^{2}} = \frac{\pi \tanh{\frac{\pi}{2} y}}{4 y}\\ 
\end{eqnarray*}
$$
\mbox{and} \qquad f' = \frac{d}{dy^2} f
$$
The manipulations needed to arrive at eq (\ref{cabkk}) are described in App \ref{manip}.

As a net result we obtain:

\begin{equation}
C^{a} + C^{b} = -\frac{1}{9} + 0.135 \epsilon^2
\end{equation}
where the leading term is the SM contribution from the top-Higgs loop with the external cross.

\subsection{Diagrams (1c) and (1d)}

Similarly to diagram (1a), we define for diagram (1c):

\begin{equation} \label{cicci}
C^{c} = C^{c}_{K0} + C^{c}_{0K} + C^{c}_{KK}
\end{equation}

It is:

\begin{equation} \begin{array}{rcl}
C^{c}_{K0} & = & \displaystyle \frac{5}{12} + \left( - \frac{\pi^2}{144} + \frac{\zeta'(2)}{6} + \frac{\pi^2}{36} \ln{\epsilon} \right) \epsilon^2 \\
 & = & \displaystyle \frac{5}{12} - \left( 0.23 + 0.27 \ln{\epsilon^{-1}} \right) \epsilon^2
\end{array}
\end{equation}

\begin{equation}
C^{c}_{0K} = \frac{5 \pi^2}{288} \epsilon^2 
\end{equation}

Unlike $C^{a}_{KK}$, however, $C^{c}_{KK}$ is finite by itself, always with the sum over KK modes made before momentum integration.
In this case the convergence occurs because of the alternating signs of the couplings $\mu_{r}$ and $\nu_{r}$ at $y = \pi R/2$.
One has:

\begin{equation}
C^{c}_{KK} = 4 \epsilon^2 \int_{0}^{\infty} 2 q^3 dq F_{c}(q) = -0.032 \epsilon^{2}
\end{equation}
where:

$$
F_{c} = \int_{q}^{\infty} 2 y dy \left( - \frac{3}{2} \left( f'_{3} \right)^2 - \frac{5}{3} f_{3} f''_{3} - \frac{2}{3} y^2 f_{3} f'''_{3} - y^2 f'_{3} f''_{3} \right)
$$
and

$$
f_{3} (y) = \sum_{n=1}^{\infty} \frac{(-1)^n}{y^2 + (2n)^{2}} = \frac{\pi y -2 \sinh{\frac{\pi}{2} y}}{4 y^2 \sinh{\frac{\pi}{2} y}}
$$

The variuos contributions in (\ref{cicci}) sum up to:

\begin{equation}
C^{c} = \frac{5}{12} - (0.09 + 0.27 \log{\epsilon^{-1}})\epsilon^2
\end{equation}
with the SM contribution reproduced in the $\epsilon \rightarrow 0$ limit.

Finally, from the diagram (1d), again without zero modes, one finds:

\begin{equation} \label{ciddi}
C^{d} = \displaystyle \sum_{n,r} C^{d}_{nr} =  4 \epsilon^2 \displaystyle \int_{0}^{\infty} 2 q^3 dq F_{d} (q) = -0.41 \epsilon^2
\end{equation}
where:

$$
F_{d} = -\frac{5}{6} \left( f'_{4} (q) \right)^2
$$

$$
f_{4} (q) = - \sum_{n=1}^{\infty} \frac{(-1)^n (2n-1)}{q^2 + (2n-1)^2} = \frac{\pi}{4} \mbox{sech}\frac{\pi}{2}q
$$

Note the exponential convergence in (\ref{ciddi}) of the integration over the momentum, rescaled by $1/R$. 
This reflects the delocalized breaking of supersymmetry, as in the case of the one loop contribution to the Higgs mass.
A similar behaviour is exibited by diagrams (1a), (1b) put together and (1c), if one takes care of not isolating the zero modes whenever they are present.

From the total of the 4 diagrams we find:

\begin{equation} \label{finres}
C = -\frac{1}{9} + \frac{5}{12} + \Delta C; \qquad \Delta C = - (0.36 - 0.27 \log{m_{t} R}) (m_{t} R)^2
\end{equation}
This is our final result.

Note that the $m_{t} R \rightarrow 0$ limit of $C$ does not reproduce the SM result even in the limit of vanishing gauge coupling.
This is because of an additional contribution $(-\frac{23}{36})$ from the diagram with the $W$ and the charmed quark as internal lines.
However, adding the KK internal states to this diagram would only introduce corrections of order $(m_{W} R\; m_{t} R)^2$.

\section{Comparison with the experiment}

Although the neglect of terms of order $(m_{W} R\; m_{t} R)^2$ or even $(m_{t} R)^4$ may be an insufficient approximation, we believe that eq (\ref{finres}) should at least give an indication of the sign and size of the physical effect of all the towers of extra states.
This allows us to examine the numerical significance of our result.

Taking into account the calculated QCD corrections, the theoretical branching ratio for $b \rightarrow s \gamma$ can be written as \cite{LOcalcolo}:

\begin{equation}
\mathcal{BR} \left. (B \rightarrow X_{s} \gamma)\right|_{th} = (3.47 \pm 0.50) \cdot 10^{-4} | 1 - 1.28 \Delta C |^{2}
\end{equation}
with our own, somewhat conservative, estimate of the error of the pure SM result.
This can be compared with the experimental branching ratio, inclusive of the entire gamma spectrum after subtraction of the intermediate $\psi$, $\psi'$ contributions \cite{CLEO}:

$$
\mathcal{BR} \left. (B \rightarrow X_{s} \gamma) \right|_{exp} = (3.22 \pm 0.40) \cdot 10^{-4}
$$

\begin{figure}[bt]
\begin{center}
\includegraphics[width=12cm]{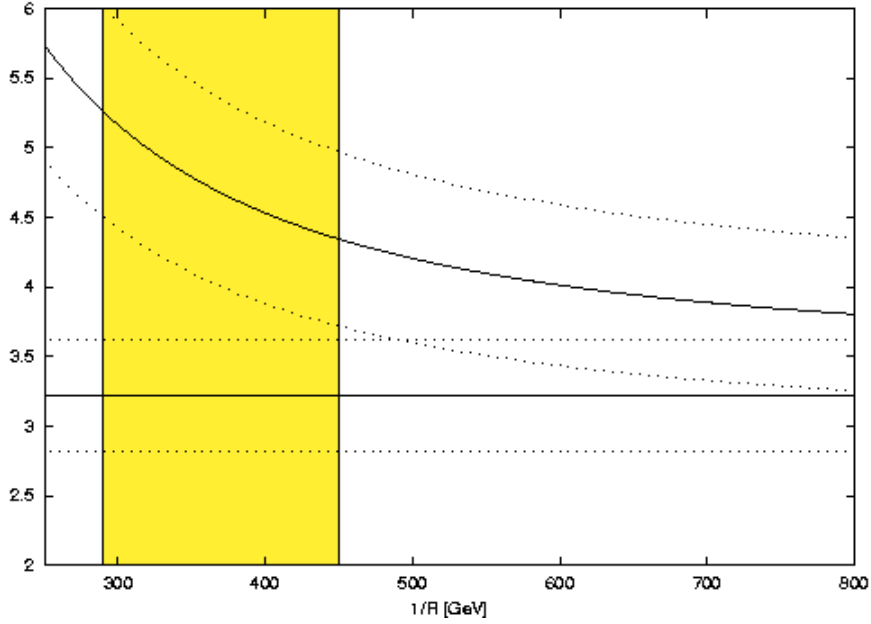}
\end{center}
\caption{$\mathcal{BR}(B \rightarrow X_{s} \gamma)$ in units of $10^{-4}$ as function of $1/R$, compared with the experimental result (horizontal band) with the relative 1$\sigma$ errors. Also shown in yellow is the expected value of $1/R$, again with its 1$\sigma$ uncertainty.} \label{comparison}
\end{figure}

The result of this comparison, with the bands of the corresponding uncertainties at $1\sigma$, is shown in Fig \ref{comparison} as function of $1/R$.
In the range of $1/R = 370 \pm 80\; GeV$ the correction due to $\Delta C$ is significant and moves the branching ratio towards higher values.
Also taking into account the approximation involved in our calculation, we think that the presence of these effects cannot, at moment, be excluded.

\appendix

\section{Diagonalization of the mass matrices.} \label{mass}

The top mass terms, generated on the $y=0$ brane, mix all the KK modes of top and stop.
In order to simplify the calculation, it is convenient to diagonalize the matrices and to work in the eigenstates basis.
This is possible because we are expanding in the parameter $\epsilon$ and it is quite simple to work with the infinite matrix at every order in $\epsilon$.

For the fermionic modes, the mass matrix can be written in the following form:

\begin{equation}
\overline{\Psi} \mathcal{M} \Psi^{c} = \frac{1}{R} \left( \psi_{T,0}, \psi_{T,l}, \psi^{c}_{U,l} \right) \left(
\begin{array}{ccc}
\epsilon & \sqrt{2} \epsilon \mathcal{I}^{T} & 0 \\
\sqrt{2} \epsilon \mathcal{I} & 2 \epsilon \mathcal{I}\mathcal{I}^{T} & M \\
0 & M & 0 \\
\end{array} \right) \left( \begin{array}{c}
\psi_{u,0} \\
\psi_{U,k} \\
\psi^{c}_{T,k}
\end{array} \right)
\end{equation}
where $M_{lk} = 2l\delta_{lk}$ is the mass matrix in the bulk, $\mathcal{I}^{T} = (1, 1, 1, ....)$ and $k, l = 1, 2, 3..$.
As the matrix is symmetric:

$$
\mathcal{M}^{diag} = V^{T} \mathcal{M} V
$$

The eigenvalue equation is:

$$
\mathcal{M} \left( \begin{array}{c}
a\\
B\\
C\\
\end{array} \right) = x \left( \begin{array}{c}
a\\
B\\
C\\
\end{array} \right)
$$
where $a$ is a number, while $B$ and $C$ are infinite-dimensional vectors.
After a straightforward algebraic manipulation, we obtain three independent relations:

\begin{eqnarray}
a & = & \sqrt{2} \frac{\epsilon}{x - \epsilon} \sum_{r} B_{r} \label{A}\\
C_{n}  & = & \frac{M_{nn}}{x} B_{n} \label{Cn}\\
B_{n} & = & 2 \frac{\epsilon x^2}{x - \epsilon} \frac{1}{x^2 - M_{nn}^{2}} \sum_{r} B_{r} \label{Bn}
\end{eqnarray}

Summing eq. (\ref{Bn}) over n, we find the exact tree-level eigenvalue condition:

\begin{equation}
\tan{\frac{\pi}{2} x} = \frac{\pi}{2} \epsilon
\end{equation}
whose solutions are \cite{BHN}:

\begin{equation} \label{eigenv} \begin{array}{rcl} 
m_{0} & = &  \displaystyle \frac{2}{\pi} \arctan{\frac{\pi}{2} \epsilon} \\
m^{\pm}_{n} & = & m_{0} \pm 2n
\end{array} \end{equation}

Then, expanding the relations (\ref{A}-\ref{Bn}) and (\ref{eigenv}), it is possible to calculate the matrix $V$ at every order in $\epsilon$, taking care of properly normalize the eigenstates.
The orthogonal matrix is easily invertible, giving the final result (\ref{psit0}-\ref{psicul}).

In the case of the stop KK modes, a similar procedure is available.
The mass matrix, relevant for our discussion, is:

\begin{equation}
\Phi^{\dagger} \mathcal{M}^{2} \Phi = \frac{1}{R^2} \left( \phi^{\dagger}_{U,l}, \phi^{c\dagger}_{T,l} \right) \left( \begin{array}{cc}
M^2 + 4\epsilon^2 k D  & -2 \epsilon D\cdot M\\
-2 \epsilon M\cdot D & M^2 \\
\end{array} \right) \left( \begin{array}{c}
\phi_{U,k}\\
\phi^{c}_{T,k}
\end{array} \right)
\end{equation}
where $M_{lk} = (2l-1) \delta_{lk}$, $D = \mathcal{I} \mathcal{I}^{T}$ and $k = \sum_{i=1}^{\infty} 1$.
A peculiarity is the presence of the divergent sum $k$, which gets cancelled in the relevant relations, as it happens in the case of the one loop corrections to the Higgs mass \cite{BHN}.

\section{Note on the calculation} \label{manip}

One needs to sum over all the internal KK mode contributions to the graphs (\ref{grapha}-\ref{graphd}) before the momentum integration to respect the relevant local supersymmetry.
In order to do that, it is best to separate the two sums, by first integrating out the Feynman parameter $x$.
We use the following formulae:

\begin{eqnarray}
\int_{0}^{1} dx \frac{x^{n-1} (1-x)^{m-1}}{[xA + (1-x)B]^{n+m}} & = & \frac{\Gamma(n) \Gamma(m)}{\Gamma(m+n)} \frac{1}{A^{n} B^{m}}\\
\frac{1}{(q^2 + a)^n} & = & n \int_{q^2}^{\infty} dy^2 \frac{1}{(y^2 + a)^{n+1}} \label{eq2}
\end{eqnarray}
where eq. (\ref{eq2}) was used to properly enhance the power of the denominator.
Then, using:

$$
\frac{1}{(q^2 + a)^{n+1}} = \frac{(-1)^{n}}{n!} \frac{\partial^{n}}{(\partial q^2)^n} \frac{1}{q^2 + a}
$$
we can write all the sums in terms of few functions, written in section \ref{calculation}.

A few words should be spent to explain why we isolated the zero mode contributions to each graph.
First of all, due to the presence of a single sum, the result for such terms is finite.
Furthermore, if this isolation were not done, the integration would became harder due to the top mass dependence of the denominators.
Letting aside these technical problems, by treating all modes on the same footing, it can be shown that the convergence at large momentum of the integrals is actually exponential for every diagram, as it is manifest in eq. (\ref{ciddi}) for graph (1d).

\section*{Acknowledgements}

This work was supported by the ESF grant under the ACROSS THE ENERGY FRONTIER contract HPRN-CT-2000-00148.

\end{document}